\documentclass[10pt, twocolumn, conference, 10pt]{IEEEtran}
\usepackage{amsfonts, amsmath, amssymb, epsfig, graphicx,subfigure}
\usepackage[linesnumbered,algoruled,boxed,lined]{algorithm2e}
\usepackage{multirow}
\usepackage{comment}
\usepackage{colortbl}

\newcommand{\textred}[1]{\textcolor{red}{#1}}
\ifx\noeditingmarks\undefined
  \newcommand{\pgwrapper}[2]{\textred{#1: #2}}
\else
  \newcommand{\pgwrapper}[2]{}
\fi

\usepackage{mathtools} \mathtoolsset{showonlyrefs}

\begin{document}

\title{Active learning for fast and slow modeling attacks on Arbiter PUFs}
\author{Vincent Dumoulin, Wenjing Rao, and Natasha Devroye \\
{\it Department of Electrical and Computer Engineering} \\
{\it University of Illinois at Chicago} \\
E-mail: vdumou2, wenjing, devroye @uic.edu}
\maketitle

\begin{abstract}

Modeling attacks, in which an adversary uses machine learning techniques to model a hardware-based Physically Unclonable Function (PUF) pose a great threat to the viability of these hardware security primitives. In most modeling attacks, a random subset of challenge-response-pairs (CRPs) are used as the labeled data for the machine learning algorithm. Here, for the arbiter-PUF, a delay based PUF which may be viewed as a linear threshold function with random weights (due to manufacturing imperfections), we investigate the role of active learning in  Support Vector Machine (SVM) learning. We focus on challenge selection to help SVM algorithm learn ``fast'' and learn ``slow''.  Our methods construct challenges rather than relying on a sample pool of challenges as in prior work. Using active learning to learn ``fast'' (less CRPs revealed, higher accuracies) may help manufacturers learn the manufactured PUFs more efficiently, or may form a more powerful attack when the attacker may query the PUF for CRPs at will.  Using active learning to select challenges from which learning is  ``slow'' (low accuracy despite a large number of revealed CRPs) may provide a basis for slowing down attackers who are limited to overhearing CRPs.
\end{abstract}

\section{Introduction}
Unclonable Functions (PUFs) are circuits with small hardware footprints and high throughputs that exploit randomness in the fabrication process to provide chip-specific, unique digital fingerprints \cite{herder2014physical}. They are hence ideally suited to form hardware security primitives to be used in a variety of settings where energy and computational resources are limited (e.g. IoT), and hence  conventional  cryptography-based security algorithms are typically too resource intensive. 

PUFs work by exploiting some inherent randomness in the manufacturing process, which produces analog features with slight random variations  that are combined to yield a   ``challenge-to-response'' Boolean function. The hope is that each device's truth table expressed as  challenge-response-pairs (CRPs) is unique and  unclonable and hence  may be used  in security applications such as device identification,
authentication, and on-demand cryptographic key generation. 

PUFs are grouped as:
1) {\it weak PUFs}, which deliver a small truth table of linear size in the number of hardware components; and
2) {\it strong PUFs}, which promise a CRP space exponential in the hardware size,  able to generate, on demand, many and/or very long keys with minimal hardware throughout a device's lifespan. 
After each chip is produced, the manufacturer either models the PUF (using machine learning) or some of the CRPs of its associated PUF will be collected in an ``enrollment'' phase. This model, or the enrolled CRPs,  will be saved on a trusted 
server to authenticate the chip in the future.
It is more attractive to learn the model, which gives a protocol the ability to request any challenge, and requires less memory for a large PUF lifetime (larger \# of CRPs).

\subsection{Contributions}
\label{sec:contributions}
In this paper, we look at how active learning, a subfield of machine learning in which one may adaptively request data points from which to learn, may be used to both speed up and slow down machine-learned modeling of arbiter PUFs.  In a ``modeling attack'' \cite{ruhrmair2010modeling}, 

 a machine learning algorithm that has access to CRPs of the unknown PUF uses these to learn a model of the PUF's input-output function with great accuracy. 
The ability to easily model an arbiter PUF may be helpful from a manufacturer's perspective, but is  undesirable from a security perspective.  

{\it Attack models and motivation.} We consider active learning in the arbiter PUF setting and assume an attacker (which may be the manufacturer itself) either a) may select which challenge (data point) to query the PUF with next, and it sees the (possibly noisy) response (label of that data point); or b)  overhears a set of CRPs which it cannot select itself, but which are nonetheless selected actively by the authenticating entity.
Attack model a) is relevant to fast learning by a manufacturer or an attacker who has access to a PUF who want to quickly learn a PUF model, and b) is relevant for helping an authentication protocol select challenges which may be overheard by an attacker, with the goal of  slowing their learning.

Active learning for APUFs has been looked at by \cite{active-PUF} (fast) and \cite{8607165} (slow), where in both cases it was  assumed the attacker has access to a  random sample pool of unlabelled data and adaptively picks from those which to next label. We look at a different setup: we do not assume access to a random sample pool but rather are able to select {\it any} challenge. Our model thus considers a sample pool the size of the entire challenge set, which for the arbiter PUF is of size $2^n$ ($n$ the PUF length) is prohibitively complex for the algorithms in \cite{active-PUF, 8607165}. We will use Support Vector Machines (SVMs) for learning. 

 {\bf Contribution 1: a PUF-realization-independent challenge set for modeling with excellent generalization error in the very small data (small \# CRPs from which to learn) regime.}

We demonstrate a set of challenges -- the Hadamard set --  that yield high test accuracy given a (small) number of data points. This set of challenges is a maximum entropy set -- meaning statistically, over all manufactured arbiter PUFs, this set should produce maximum entropy responses. 
 
{\bf Contribution 2: challenge construction for ``fast'' active learning without a sample pool.} 
    rather than rely on a sample pool of challenges, we construct  the next challenge according to a desired approximate distance to the learned hyperplane while ensuring it is uncorrelated with previously constructed challenges. Our proposed algorithm has a tunable parameter controlling the distance to the hyperplane: taking this close to 0 yields challenges which are likely to yield new support vectors.
    
    This is the fastest  arbiter PUF modeling attack  we are aware of.
 
  {\bf Contribution 3: challenge construction for ``slow'' active learning without a sample pool.} By picking the tunable parameter differently, we can ``slow'' down learning when the attacker only has access to the selected overheard challenges. We  select challenges which will result in poor accuracy:  10,000 challenges can be overheard yet still lead to an accuracy of around  70\%.  This is based on selecting challenges which are {\it not} too close to the hyperplane (not too informative, no new support vectors), but also not too far (which would yield correlated challenges as so few are very far).

\section{Arbiter PUF}

We first provide an analytical model for  the input-output function of the 
Arbiter PUF (APUF) \cite{gassend2004identification}, viewed as a linear threshold function.
A challenge vector 
${{\bf \Phi}}:= (\phi_1, \phi_2, \cdots \phi_{n+1}) \in \{\pm 1\}^{n+1}$, where $\phi_{n+1}=+1$ is a vector that is input to the PUF, and which responds with a one-bit response $ R_{\bf \Phi}$:

\begin{align}
 R_{\bf \Phi} = \text{sign}({\bf \Phi} \cdot {\bf w}) = \text{sign}\left( \sum_{i=1}^{n+1} \phi_i w_i\right) \in \{\pm 1 \}, \,
 \label{eq:arbiter2}
\end{align} 
where
  ${\bf w}:= (w_1, w_2, \cdots w_{n+1}) \in \mathbb{R}^{n+1}$ is a vector depending solely on the random manufacturing delays, and $\cdot$ denotes the inner product of two vectors,  $\text{sign}$ is the signum function which takes the sign of the argument, and assigns it $\pm1$ with probability $\frac{1}{2}$ if the argument is 0 (happens with measure 0). These $w_i$'s are different for each manufactured PUF instance, and are often assumed to follow a zero-mean Gaussian distribution \cite{boning2000models}.
This representation of a PUF eases the analysis  as the output is now expressed as a {\it  linear threshold function}: one can then visualize the PUF Boolean input-output function as splitting a high dimensional hypercube with a hyperplane: all vertices on one side are are classified are $-1$, while the others are classified as $+1$.

{\bf Challenge correlation:} Prior work \cite{us-TCAD} has found the exact probability, taken over the randomness in the PUF manufacturing process (the random $w_i$'s), that two arbitrary challenges produce the same response, which depends on the challenge correlation, defined as $\Gamma/n: = (|{\cal S}| - |{\cal F}|)/n$ for ${\cal S} := \{i \in \{1, \cdots n\}: \phi_i = \phi'_i\}$ (the number of ``same'' elements in the two challenges) and ${\cal F}:= \{i \in \{1, \cdots n\}: \phi_i \neq \phi'_i\} $ (the number of ``flipped'' elements in the two challenges).  We will call two challenges ${\bf \Phi}$ and ${\bf \Phi'}$ {\it uncorrelated} if $ P[R_{\bf \Phi} = R_{{\bf \Phi'}}] = 0.5$. The authors showed \cite{us-TCAD} this happens when $\Gamma = 0$ when $\phi_1=\phi_1'$ (i.e. $|{\cal S}| = |{\cal F}|$),  or when  $\Gamma +1 =0$ (i.e. $|{\cal S}| = |{\cal F}| - 1$) when $\phi_1\neq \phi_1'$.   It was also shown that the number of  challenges that are uncorrelated (with a given challenge) make up the vast majority of challenges. Thus, if challenges are randomly generated, ``most'' are uncorrelated.

\section{Modeling attacks on Arbiter PUFs}

The fact that arbiter PUFs may be represented as linear threshold functions is one reason why they have been so easy to model \cite{ruhrmair2010modeling, ruhrmair2013puf, 7096998}. 

Learning linear threshold functions is particularly well suited to SVM learning, which we briefly outline next. 

\subsection{Support vector machines}
Support vector machines (SVM) are supervised learning models with associated learning algorithms that analyze data for classification and regression analysis \cite{Vapnik1995}.  Assume we are given a training dataset of labeled data, which in the case of an APUF corresponds to $({\bf \Phi}, R_{\bf \Phi})$ challenge-response pairs. 
\begin{figure}
    \centering
\includegraphics[width=0.4\textwidth]{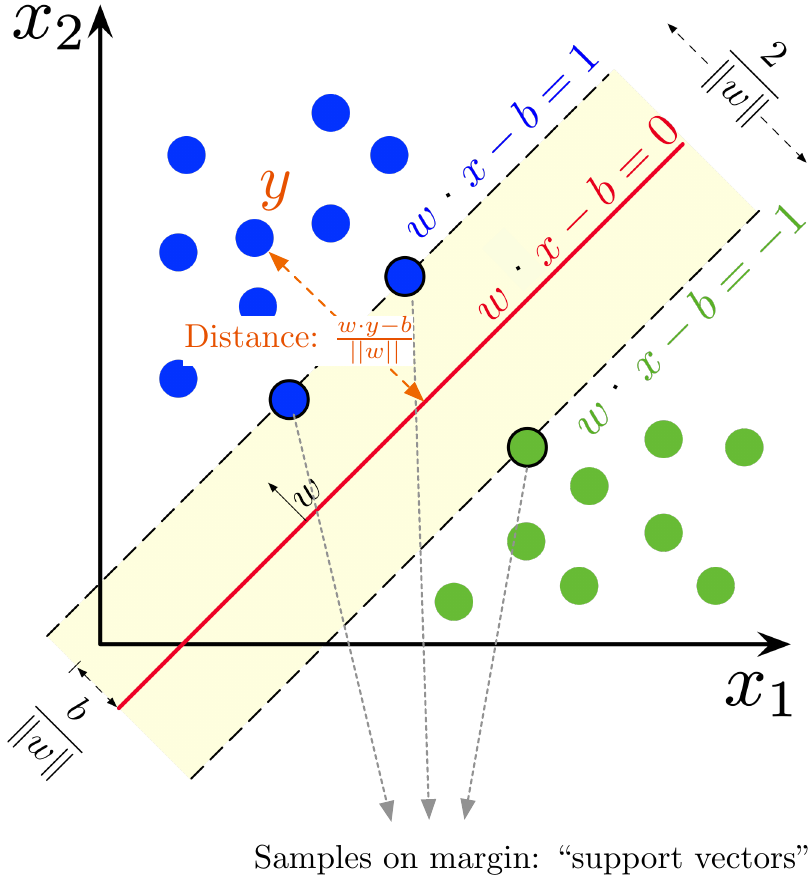}
    \caption{Illustration of an SVM, adapted from \cite{SVM-wiki} showing ``support vectors'' and the hyperplane described by $w\cdot x-b=0$ separating two classes of data (blue and green). The distance of orange data point $y$ from the hyperplane is given by $w\cdot y-b$.}
    \label{fig:SVM}
\end{figure}

In linear SVM, one aims to find the ``maximum-margin hyperplane'' that divides the group of challenges with $R_{\bf \Phi}=1$ from the group of challenges with $R_{\bf \Phi}=-1$. 
A hyperplane can be written as the set of points ${\bf x}\in \mathbb{R}^n$ satisfying:
\begin{equation}\label{eq:hyperplane}
    {\bf w}\cdot{\bf x}-b=0 \;\; \text{ for some ${\bf w}\in \mathbb{R}^n$ and $b\in \mathbb{R}$.}
\end{equation}
For the APUF, we know that the data or challenges (in the absence of noisy labels) may be exactly linearly separated by a hyperplane of weights ${\bf w}:= (w_1, w_2, \cdots w_{n}, w_{n+1})$, with $b=0$ (recalling that $\phi_{n+1}=1$ for all challenges).  Thus, using SVM in a modeling attack is sensible as the model matches the physical model of an APUF.  Fig. \ref{fig:SVM} illustrates SVM  in which the data consists of two-dimensional vectors in $\mathbb{R}^2$, each with a label (blue or green), separated by the red learned hyperplane.  
If ${\bf y}$ is a data point (to be thought of as an $n+1$-dimensional challenge, with the color indicating the response of the PUF to that challenge), then the distance to the hyperplane  ${\bf w}\cdot {\bf x} -b=0$ is given by $({\bf w}\cdot {\bf y} - b)/||{\bf w}||$ for $||{\bf w}||$ the norm of the vector ${\bf w}$. The vectors that lie closest to the hyperplane are called the support vectors. For APUFs, the hyperplane goes through 0, i.e. $b=0$.

\subsection{Active learning using SVMs}

Active learning using SVMs has been considered, see for example \cite{rep-inform, Kremer}. When selecting a next data point to learn from, a tradeoff between representativeness (selecting data which represents the true distribution from which the data is sampled, which may sometimes lead to highly correlated samples) and informativeness (selecting data which reduces generalization error by providing ``new'' or ``informative'' data points) often exists. In the context of the APUF, we may understand this tradeoff intuitively as follows:

\noindent $\bullet$ Representativeness: implies that the data points, which are challenges, are selected according to the true distribution. In this case, the representative distribution is random, with all challenge vectors being equally likely. From \cite{us-TCAD} we know that it is highly likely that two randomly selected challenges are uncorrelated, or $|{\cal S}| \approx |{\cal F}|$.  So, we will use $\Gamma := |{\cal S}| - |{\cal F}|$ as a measure of representativeness, with $\Gamma\approx 0$  meaning the data is more representative.

\noindent $\bullet$ Informativeness: in the context of SVM implies that we select challenges which are likely to be new support vectors, i.e. decrease the margin, or lie close to the hyperplane. This challenge will depend on our current model of the APUF, where by model we mean the estimated weights ${\bf w} = (w_1, w_2, \cdots w_n, w_{n+1})$. In that case, one  aims to select a challenge ${\bf \Phi}$  close to the hyperplane, meaning {$\frac{{\bf w}\cdot {\bf \Phi}}{||{\bf w}||} \approx 0$.}

\subsection{Modeling attacks on arbiter PUFs}

{\bf Passive modeling attacks on arbiter PUFs.} 
Most existing modeling attacks on APUFs \cite{ruhrmair2010modeling}  assume the attacker has access to a {\it randomly selected} set of challenges which they use to train their models using logistic regression, SVM, and evolutionary strategies. 
 This is a passive type of learning -- the challenges selected do not depend on the model. 
 
 Active learning, where a PUF manufacturer or attacker  actively queries the APUF with adaptively selected challenges, has the potential to accelerate (or decelerate) the learning. 
We are aware of only two papers which applied active learning to PUFs:  \cite{active-PUF} for ``fast'' active learning, and \cite{8607165} for ``slow'' active learning.

{\bf Active learning-based modeling attacks on arbiter PUFs: fast learning.} 
In \cite{active-PUF}, a pool of challenges was assumed to be given, and the adversary adaptively  selects  (meaning select and query a challenge, receive the response, update the model and repeat) which challenge to query based on either: 

1) uncertainty sampling: all challenges in the pool are ordered according to their ``uncertainty'' score defined as $|P[R_{\bf \Phi} = +1 |{\bf \Phi}] - P[R_{\bf \Phi} = -1|{\bf \Phi}]|$. The one with the minimal score, i.e. in which the likelihood of yielding positive and negative responses is closest, is selected. One issue is how to obtain accurate estimates of the $P[R_{\bf \Phi} = +1 |{\bf \Phi}]$ and $P[R_{\bf \Phi} = -1 |{\bf \Phi}]$, especially for the first few rounds with inaccurate  models. 

2) estimated error-rate as uncertainty sampling:  the challenges in the pool are ranked according to their estimated delay differences $|\Delta_n({\bf c})|=  |{\bf \Phi}\cdot {\bf w}|$. Select the one with the smallest value, i.e. close to the arbiter's threshold of 0. The value $|\Delta_n({\bf c})| = |{\bf \Phi}\cdot {\bf w}|$ is obtained using the current model of the PUF.

3) query-by-committee method: various models of the same labeled set are kept which represent competing hypotheses. 

The challenge where the committee members disagree on most is selected.

{\bf Active learning-based modeling attacks on arbiter PUFs: slow / adversarial learning.} 
 In the slow learning case,  \cite{8607165} looks at CRP selection to obtain a set of challenges  which result in very high internal recognition (validating or testing the model on challenges within the set but not used for training yields accuracy around 99-100\%) but poor external recognition / generalization error (validating or testing the model on challenges outside the set yields poor performance). The authors use AdaBoost to order a very large set of challenges according to AdaBoost's perceived ability of each challenge to aid in modeling the APUF. Through this ordering as a by-product of AdaBoost, they select challenges which appear to slow learning. This requires a large amount of computation and is experimentally driven. It is motivated by man-in-the-middle attacks where an attacker intercepts challenges sent by the server and their responses from the chip/user, and stores this information to build a model of the PUF in order to clone it. Having high internal recognition fools the attacker into thinking it has accurately cloned it, whereas it has not if the external recognition / generalization error is poor.

\section{Learning without a sample pool: small challenge set fast learning}

{\bf Goal:} find a set of challenges which form a good (lead to high accuracy models)  ``initial'' small set of challenges to use when using machine learning to model arbiter PUFs.

We first consider the case in which we build a model from less than $n = \# $ arbiter PUF stages number of CRPs. This is quite a small number of CRPs by modeling attack standards.  For any $k<n$ we propose to use a set of $k$ challenges, independent of the PUF realization (and hence not active learning per-se, just a specific challenge set selection), which forms part of the  binary Hadamard code of length $n$ as the challenges from which to learn. We hypothesize  that this is the set which is {\it independent of the PUF realization} which will yield the smallest generalization error. As such, if we only have a small number of challenges available, we hypothesize that this is the most desirable set of challenges to learn from.

In order to understand this set and its choice, note that in \cite{rioul2016entropy} it was shown that for a loop PUF, up to equivalence, the set of $n$ challenges that maximizes the entropy of the bit vector of $n$ responses to those challenges are the rows of the Hadamard matrix of size $n\times n$\footnote{The authors \cite{rioul2016entropy} have results for $>n$ challenges but we will not delve into those here. For $<n$ challenges, simply taking a subset of the Hadamard matrix maximizes the output-bit-vector entropy.}, yielding the maximal entropy value of $n$ bits. 
A Hadamard matrix is a square matrix whose entries are either $+1$ or $-1$ and whose rows are mutually orthogonal, meaning their inner product is 0.

Hadamard matrices are of order $2^k$ where $k$ is a non-negative integer.

We propose to use the Hadamard ``equivalent'' set of challenges for the arbiter PUF in order to maximize the response entropy to a value of $n$. Intuitively, one desires maximal entropy in the responses in order to ensure that these challenges produce maximally informative outputs from which to learn. 
All the aforementioned results were presented for the loop-PUF, which is a delay-based simplification of the arbiter PUF.  The entropy-maximizing challenges for a loop PUF do not translate directly to an arbiter PUF, but may be done with some care.

{\bf Hadamard-like set for fast learning.} To guarantee that any pair of challenges ${\bf \Phi}, {\bf \Phi'}$ are orthogonal is equivalent to trying to guarantee that $\Gamma=0$ (if $\phi_1=\phi_1'$) or $\Gamma+1=0$ (if $\phi_1\neq\phi_1'$). %
This may be done by constructing an $n\times n$ Hadamard matrix using Sylvester's construction \cite{Sylvester1867} which ensures the first row is all ones, and all pairs of rows mutually orthogonal. Taking the challenges as the rows of this matrix, appended with an all ones column (so rows are of dimension $n+1$ as needed, and $\phi_{n+1}=1$) we will ensure that the challenges produce pairwise uncorrelated outputs and hence are likely a good choice for initial learning of PUFs, which we experimentally validate. 

\begin{figure}[h]
    \centering
\includegraphics[width=0.49\textwidth]{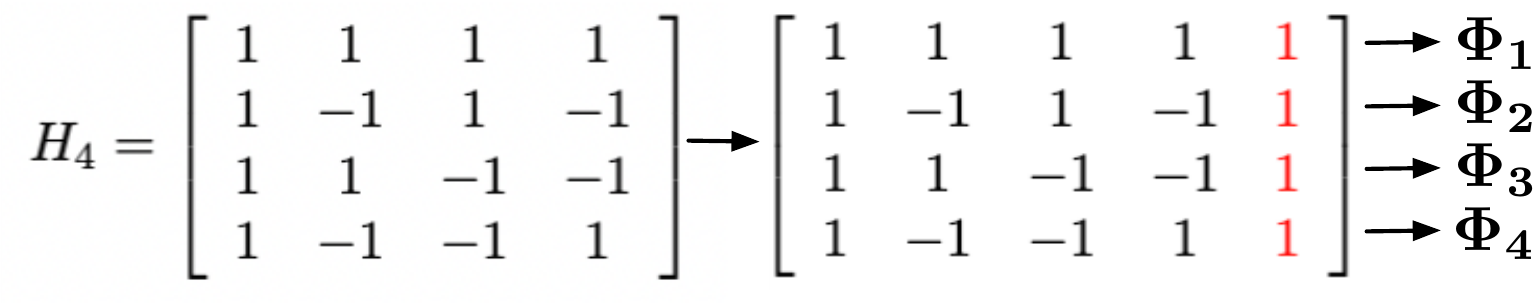}
    \caption{Example creation of a set of $n=4$ challenges for a length $n=4$ PUF that form a Hadamard set and are hence good for ``fast'' modeling with low \# of challenges.}
    \label{fig:Had}
\end{figure}

{\bf Example Hadamard-based challenge set creation for $n=4$.} As an example, we consider an APUF with $n=4$ stages. We aim to create a set of 4 challenges which are statistically orthogonal and hence are a good choice from which to learn ``fast''. To do so, we look up the Hadamard matrix of size 4 constructed using Sylvester's method \cite{Sylvester1867}, as $H_4$, append a column of 1's to give us 4 length 5 vectors ${\bf \Phi_1, \cdots \Phi_4}$ which are our challenges, all as shown in Figure \ref{fig:Had}.

\begin{figure}[h!]
	\centering	\includegraphics[width=0.5\textwidth]{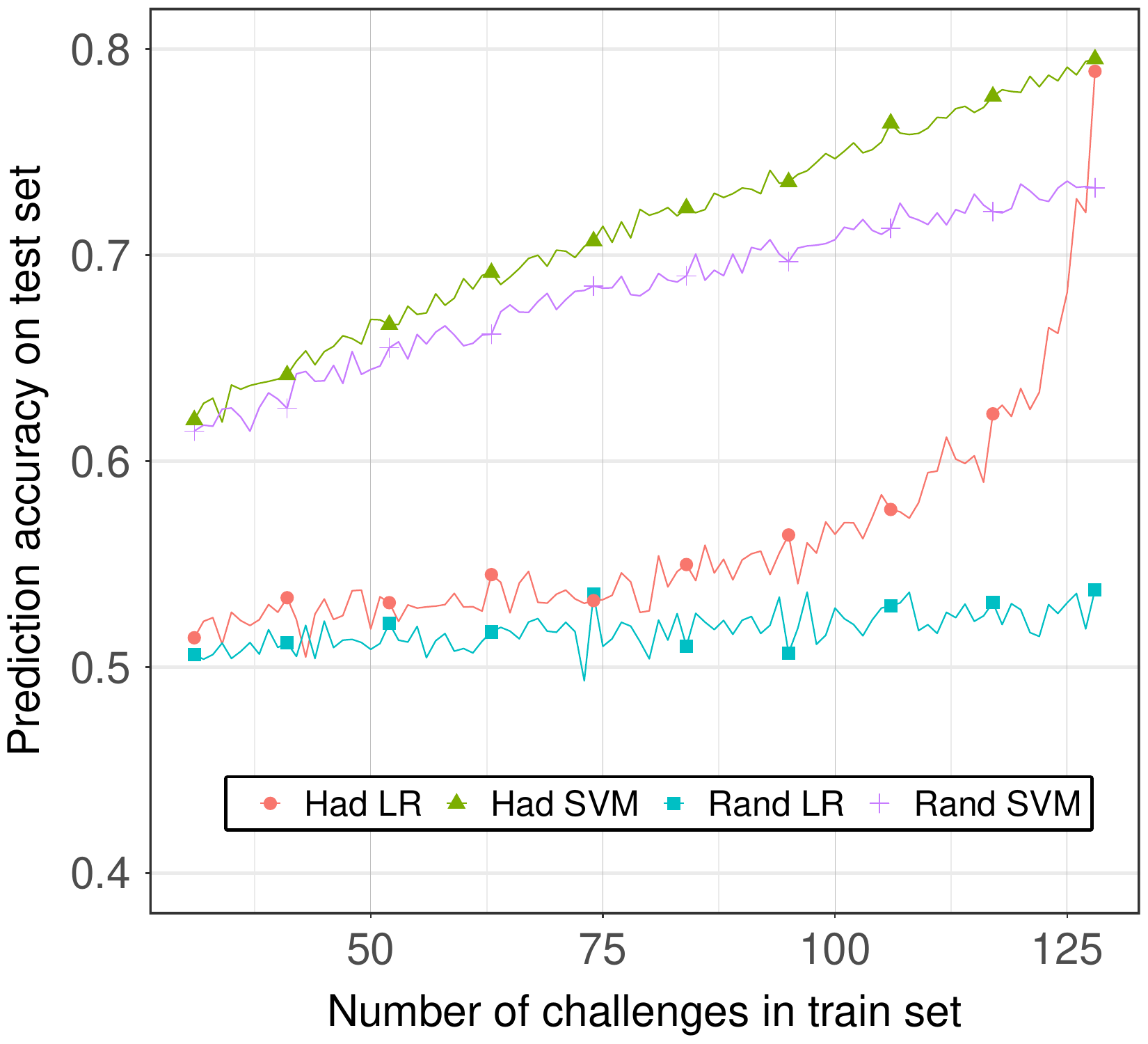}
		\caption{Hadamard set accuracy arbiter PUF 128 stages}		\label{fig:hadamard_acc}
\end{figure}

Figure \ref{fig:hadamard_acc} shows the prediction accuracy (percentage of correctly predicted responses) on test sets (randomly generated challenges) of 4 learning algorithms for an arbiter PUF with $n=128$ stages. ``Rand'' means challenges in the sets used for learning were generated randomly, while ``Had'' means they were selected from the Hadamard set for $n=128$. The x-axis plots the number of challenges used by either the Support Vector Machine (SVM) or Logistic Regression (LR) modeling algorithms. Notice we plot the prediction accuracy for every integer number of (challenge, response) pairs (CRPs) these algorithms learn from to compare their progression using randomly generated challenges versus ones selected from the Hadamard set. Notice that at such extremely low number of data points, or revealed CRPs, Hadamard-based SVM have the best performance, but around $n=128$ LR using the Hadamard set catches up. We cannot yet explain why LR with the Hadamard set suddenly catches up, while under random learning LR takes much longer to catch up other than the  intuition that the Hadamard set provides a maximal entropy set and hence is likely to provide a  good initial set of data points from which to learn for most learning  algorithms.

\begin{figure}[h!]
	\centering	\includegraphics[width=0.5\textwidth]{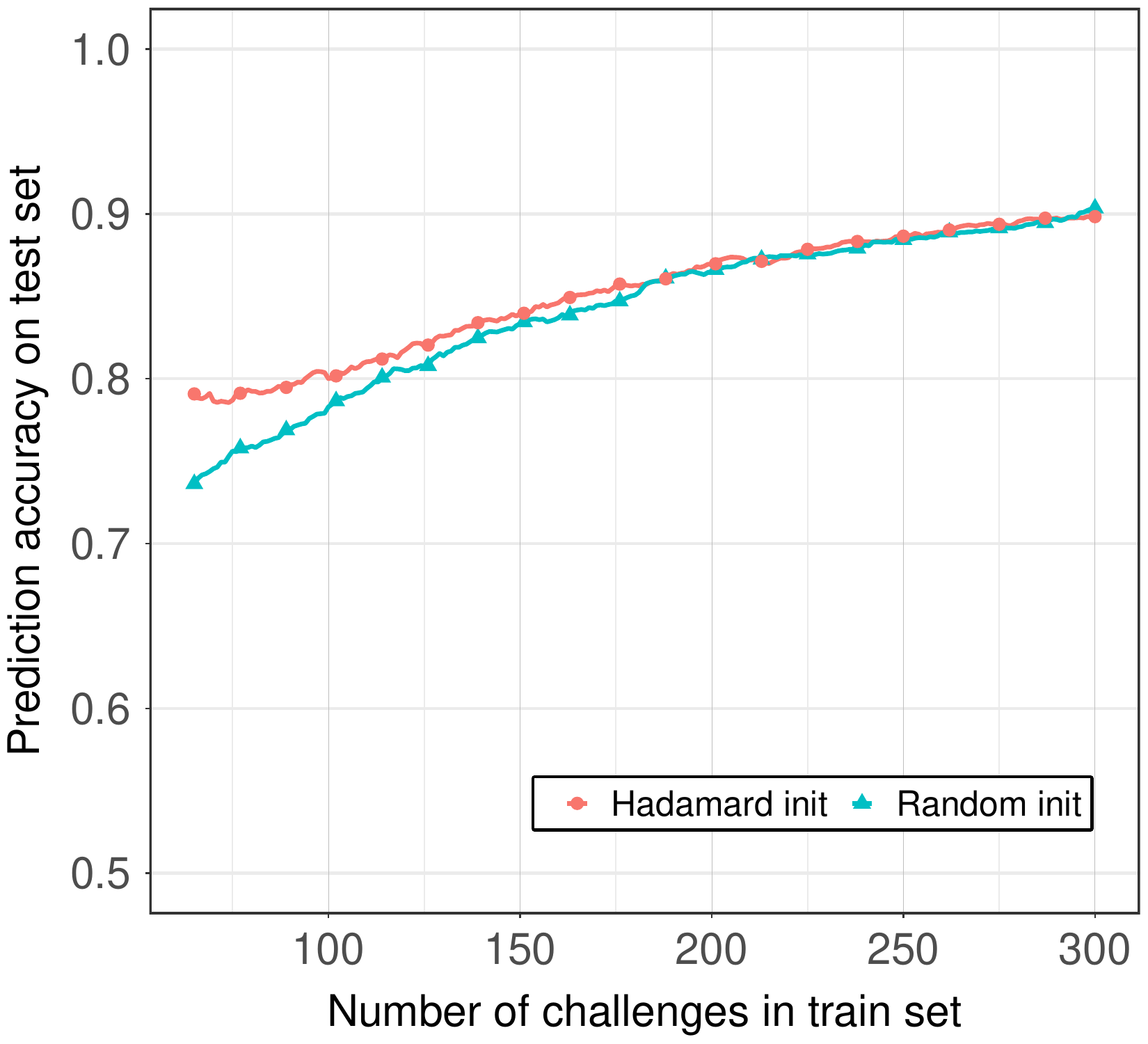}
		\caption{Hadamard vs random initialisation, PUF 64 stages.}		\label{fig:rand_vs_had_init}
\end{figure}

Figure \ref{fig:rand_vs_had_init} extrapolates the red ``Had SVM'' and blue ``Rand SVM'' lines beyond $n=128$ challenges (the limit of Hadamard-based challenge selection presented here). Beyond 128 challenge, both learn from randomly selected challenges. We see that while the Hadamard set is very good for a low number of challenges, randomly selected challenges soon catch up (another 100 or so challenges for the $n=128$ PUF and both have similar performance). As such, which to use depends on the application and whether we seek to learn from $<n$ challenges (in which case SVM with Hadamard challenges is the way to go) or more (in which case SVM with random or Hadamard initialization, random after, both work well).

\begin{figure*}[h!]
	\centering	\includegraphics[width=\textwidth]{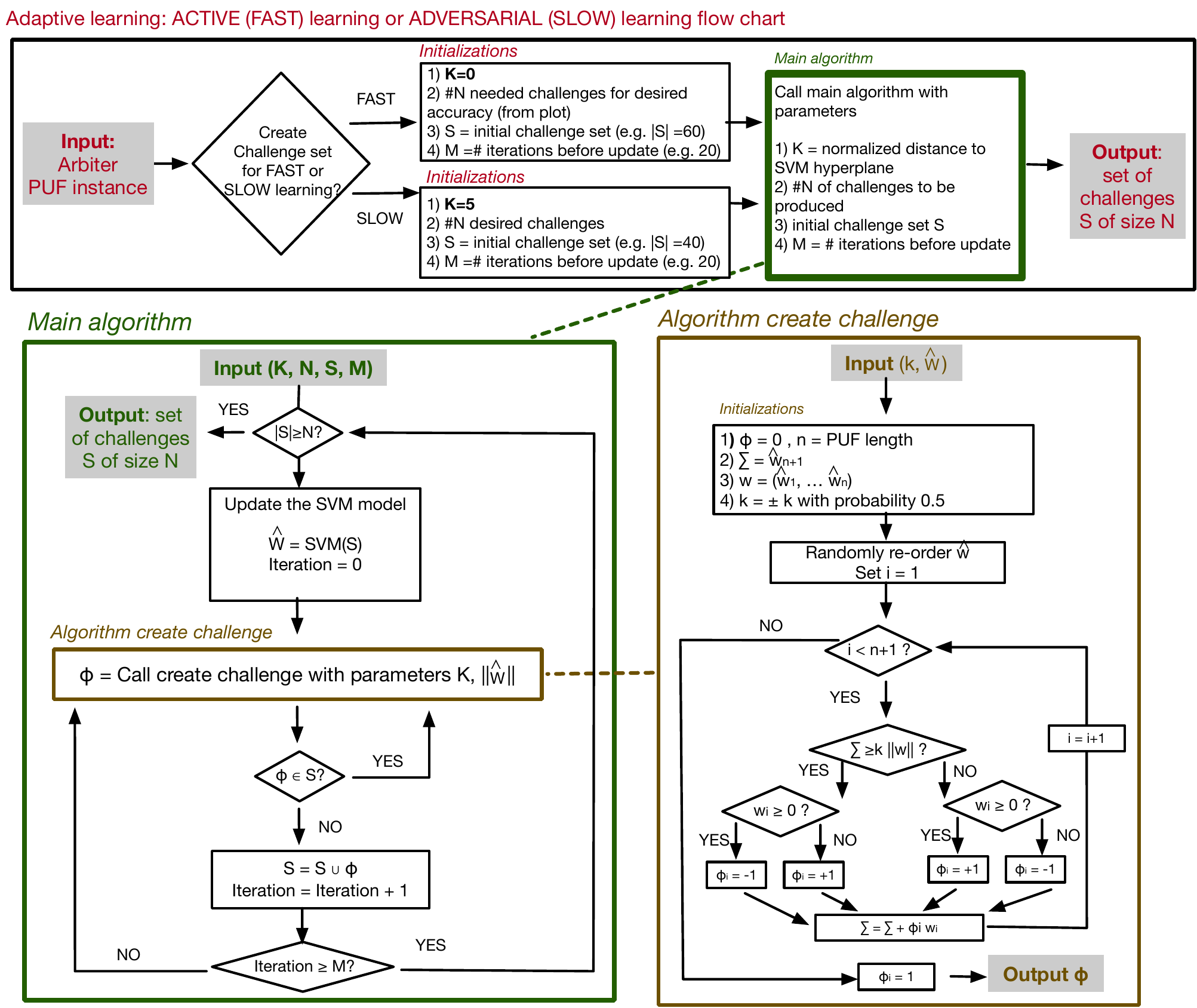}
		\caption{Flow diagram of our proposed active learning for both fast and slow learning.}		\label{fig:algorithms}
\end{figure*}

\section{Adaptive learning without a sample pool: fast and slow learning}

{\bf Goal:} We now aim to design a simple active learning technique (which may depend on the actual PUF realization) which selects the next challenge or batch of M challenges without the use of a challenge pool. For attack models and motivation, see Section \ref{sec:contributions}.   This is markedly different from prior work in active learning-based modeling attacks where sample pools are assumed, which, for large sample pools, 
can become computationally timely and/or expensive. Here, there is no sample pool and we construct the next  challenge directly. One can thus access the huge challenge space of size $2^n$, where $n$ is the PUF-length. We are not aware of any prior work on active learning without a sample pool. The model can be updated every CRP or after adding say $M = \#$ challenges. This parameter $M$ trades-off performance and computational time. In our experiments taking $M=20$ yielded good performance.

 {\bf Algorithm ``adaptive  learning''} (Fig. \ref{fig:algorithms} top flow, in red) takes as input a set of initial challenges ${\cal S}$ (to learn a rough initial model of the APUF, this set may be Hadamard or random), and a parameter $k\in \mathbb{R}$, termed the ``normalized distance'' parameter\footnote{Strictly $k$ is not a normalized distance since it can be negative. One can interpret this as follows: positive $k$ values are a certain distance on one side of the hyperplane, and negative ones on the other side of the hyperplane.} which may be tuned by the modeler (the attacker, or the manufacturer) to attain varying learning speeds. It will call our main algorithm once the initial parameters are set up.

 {\bf Algorithm ``main algorithm'':} (Fig. \ref{fig:algorithms} in green) is based on using SVM-specific information: it seeks challenges that lie at different desired distances to the hyperplane (different levels of ``informativeness''), yet are uncorrelated with previously queried challenges (hence are ``representative'' since most real challenges are uncorrelated \cite{us-TCAD}). 
 Essentially we will seek to select challenges at a distance approximately $\pm k||{\bf w}||$ to our current estimate of the hyperplane ${\bf w}\cdot {\bf x} = 0$, for ${\bf x} \in \{\pm 1\}^{n+1}$ to be thought of as the challenge ${\bf \Phi}$. Note that this $\pm k$ means that with probability half we select $+k$ and with probability half we select $-k$.

\noindent $\bullet$ For $k\approx 0$ we obtain fast learning. To intuitively see why, taking $k\approx 0$ selects challenges that lie close to the current model of the separating hyperplane. This means it will select challenges that lie close to the decision boundary and hence can help to catch the subtle differences and are likely to create new support vectors in the SVM learning algorithm.

 \noindent $\bullet$ 
 Interestingly. as $k$ increases, the learning speed slows down. We view this algorithm with tunable $k$ parameter as being potentially useful in scenarios where different learning speed are desired. As we will see experimentally in the next section, challenges that are far from the hyperplane ($k\in [3,5]$ say), but not too far lead to poor learning. Once you go beyond that for large $k$,  there are few statistically uncorrelated challenges: this is easy to intuitively understand as ${\bf w}\cdot {\bf \Phi} = \Delta_n({\bf \Phi})$ and it is hard to get very many large or small values of this -- one would need all the delay stages to align such that they add constructively.

 To use this algorithm for ``fast'' learning, we first need to run it for various number of challenges to produce the graph as in Figures \ref{fig:SVMnonoise} and \ref{fig:SVMnoise} so we know approximately how many challenges $N$ need to be used to attain a desired accuracy of say $A$. For ``slow'' or adversarial learning we can simply select the number of challenges we would like to create, ignoring the desired accuracy (ideally as low as possible).

{\bf Algorithm ``create challenge'':} (Fig.\ref{fig:algorithms} in brown) called by our main algorithm to create a challenge ${\bf \Phi}\in \{\pm 1\}^{n+1}$ that is statistically uncorrelated with others created  and yields a desired normalized distance to a given hyperplane described by ${\bf w}\in \mathbb{R}^{n+1}$. It does so by randomly permuting the indices and then selecting each $\phi_i$ value so that $\sum_{i=1}^{n+1} \phi_i w_i \approx k ||{\bf w}||$. 

This random permutation is designed to produce challenges that are generally statistically uncorrelated.

{\bf Remark:} One could ask whether this active learning approach extends to non-linear extensions of the arbiter PUF, e.g. the XOR, feedforward, or iPUFs.  For example, XOR PUFs can be represented as a threshold function with a non-linear boundary \cite[Eq. (4)]{ruhrmair2010modeling} with a small number of parameters, or with a linear boundary \cite[Eq. (5)]{ruhrmair2010modeling} with a much larger number of parameters. Our method's geometric interpretation requires a linear threshold, so one could in theory apply it to XOR's large dimension linearization. However we expect that to learn this large dimension, even with active learning, would require many more challenges than simply learning the non-linear boundary as done in \cite{ruhrmair2010modeling}. Extending our active learning technique to non-linear threshold functions is an interesting open problem.

\section{Numerical results}

We now present numerical validations of the proposed Active learning algorithm, both for fast and slow learning. 
We consider both noiseless and noisy ($Var_{intra}=3.5\%$ resulting in  approximately 3.5\% of CRPs' responses flipping) data samples. We look at either the prediction error rate  (\% of responses that were incorrectly predicted by the learned model on a randomly generated set of test challenges), or the accuracy (\% of correctly predicted by the learned model on a randomly generated set of test challenges). Both are averaged over 50 PUF instances to reduce the variance.
All our learning algorithms are based on SVM for $n=64$ length arbiter PUFs unless explicitly noted. The number of challenges before updating the model $M=20$.
The proposed method was implemented in R, using the SVM function from the package ``e1071''.

{\bf Active / fast learning.} Table \ref{table:us-nonoise} compares the performance of passive sampling of  {\it random} data points / challenges (the label random in the table) to our proposed {\it active} learning algorithm run with parameter $k=0$ (sampling challenges that lie close to the current model of the hyperplane) in the absence of noise (no noisy CRPs). The initial set $S$ used to train the first model contains $60$ random challenges. We see extraordinary improvement of our method over random sampling under SVM learning, achieving just under 97\% accuracy with a mere 350 challenges for an $n=64$ arbiter PUF. This is the fastest to date presented learning algorithm for arbiter PUFs we are aware of.

From Tables \ref{table:others-noise} (taken from \cite{active-PUF} directly) and Table \ref{table:us-noise}, in 3.5\% noise,  to obtain a prediction error of 5\% ($Var_{intra} = 3.5\%$), more than 1000 CRPs are required in passive learning with randomly selected challenges, 811 are needed by the active learning algorithm of \cite{active-PUF}, while only 300 CRPs are required when using our proposed active learning method with $k=0$. This highlights the strength of our approach: since we are not limited to a sample pool, we can more effectively select challenges from which to learn. In addition, our algorithm is remarkably simple, making smart use of the linear threshold model for the arbiter PUF and its suitability to SVM learning.

\begin{table}[!h]
    \centering
\begin{tabular}{ |c|c|c|c|c|c| }
 \hline
 \multirow{2}{*}{Sampling strategy} & \multicolumn{5}{|c|}{Number CRPs in training set} \\
 \cline{2-6}
 &$200$&$350$ & $550$ & $750$ & $1000$\\
 \hline
 Random &$14.6$&$8.8$&$5.6$&$4.0$&$3.2$\\\hline
 Active &$9.0$&$3.0$&$1.4$&$1.00$&$0.6$\\\hline
\end{tabular}
\caption{Active / fast learning, no noise, prediction error (\%)  using Algorithm ``Active learning'' with $k=0$}
\label{table:us-nonoise}
\end{table}


\begin{table}[!h]
    \centering
\begin{tabular}{ |c|c|c|c|c|c| }
 \hline
 \multirow{2}{*}{Sampling strategy} & \multicolumn{5}{|c|}{Number CRPs in training set} \\
 \cline{2-6}
 &$350$&$550$ & $1250$ & $1750$ & $6000$\\
 \hline
 Random &$11.5$&$8.10$&$6.12$&$5.90$&$4.30$\\\hline
 EQB &$9.35$&$5.90$&$4.75$&$4.33$&$4.20$\\\hline
 Error rate-based &$7.28$&$5.52$&$4.07$&$3.43$&$2.70$\\
 \hline
\end{tabular}
\caption{Active / fast learning, work by \cite{active-PUF}: $3.5\%$ noise, prediction error $(\%)$}
\label{table:others-noise}
\end{table}


\begin{table}[!h]
    \centering
\begin{tabular}{ |c|c|c|c|c|c| }
 \hline
 \multirow{2}{*}{Sampling strategy} & \multicolumn{5}{|c|}{Number CRPs in training set} \\
 \cline{2-6}
 &$200$&$350$ & $550$ & $750$ & $1000$\\
 \hline
 Random &$15.9$&$10.0$&$7.37$&$6.46$&$5.6$\\\hline
 Active &$10.4$&$5.4$&$3.9$&$3.6$&$3.5$\\\hline
\end{tabular}
\caption{Active / fast learning, $3.5\%$ noise, prediction error (\%) using Algorithm ``Active learning'' with $k=0$}
\label{table:us-noise}
\end{table}

{\bf Adversarial / slow learning.} 
Now we look at slow learning, and make the assumption that all overhead CRPs are accurate (no noise) which is the worst case for us; noise only makes the PUF harder to learn (in this case low accuracy is desired). We use Algorithm ``Active learning'' with $k=5$ and a initial set $S$ such that $|S|=40$ to generate a set of 10000 new challenges that yields poor generalization error.

To verify that this set does  not effectively learn the PUF we use it to train various state-of-the-art learning algorithms: LR, SVM, Adaboost (with a decision tree as base-learning method) and artificial neural networks with a sigmoid activation function. 
AdaBoost (Adaptive Boosting) is a classification method that combines weak / simple learners (but can also combine strong learners) to build a stronger learner.  AdaBoost seems to perform well when the individual learners consist of decision trees \cite{adaboost_tree:2013}.

For the first two, training a model on the first 500 challenges gives an accuracy of 100\% on the other challenges within the set. For the two last ones, 10-fold cross validation is used to determine the tuning parameters: number of trees (50) and max depth (1-3); number of hidden layers (1-3) and weight decay (0.1). Cross-validation yields the optimal parameters with an internal recognition of 100\%. The external recognition or generalization error is bad for all and lower than 70\%, meaning that the attacker, while it might believe it has learned the PUF model due to the high internal recognition,  failed in accurately cloning the PUF.

\begin{table}[!h]
    \centering
\begin{tabular}{ |c|c|c|c|c| }
 \hline
 \multirow{2}{*}{ML method} & \multicolumn{4}{|c|}{Number CRPs in training set} \\
 \cline{2-5}
 &$1000$&$3000$ & $5000$& $10000$\\
 \hline
 LR &$63$&$63$&$63$&$64$\\\hline
 SVM &$67$&$68$&$68$&$68$\\\hline
 Nnet &$67$&$68$&$68$&$68$\\
 \hline
  Adaboost &$66$&$67$&$67$&$67$\\
 \hline
\end{tabular}
\caption{Adversarial / slow learning, no noise, accuracy (\%)  using Algorithm ``Active learning'' with $k=5$, $|S|=40$.}
\label{table:adversarial}
\end{table}

\begin{figure}[h!]
	\centering	\includegraphics[width=0.45\textwidth]{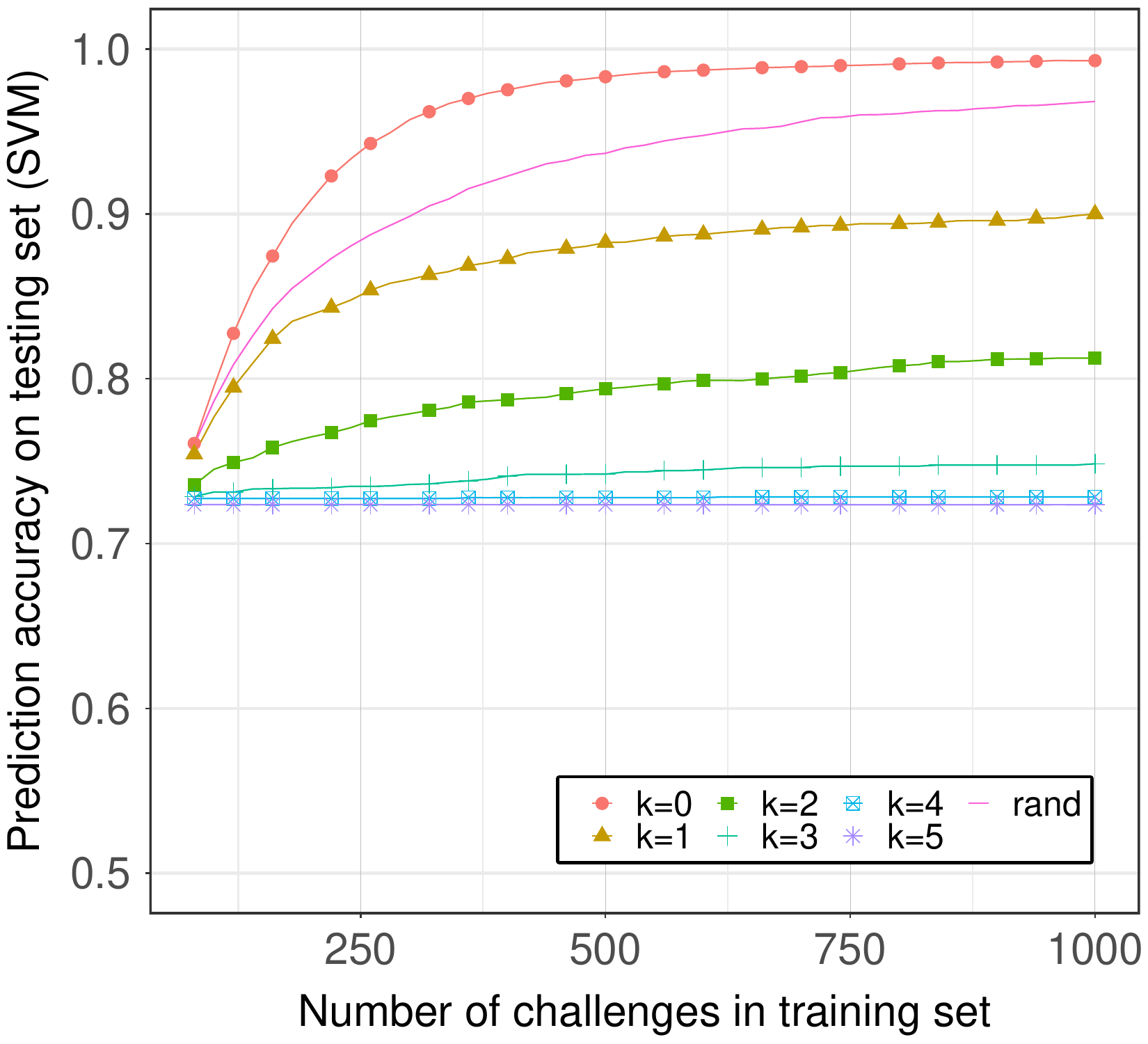}
		\caption{No noise, accuracy, SVM, for a $n=64$ stage APUF}		
  \label{fig:SVMnonoise}
\end{figure}

\begin{figure}[h!]
	\centering	\includegraphics[width=0.45\textwidth]{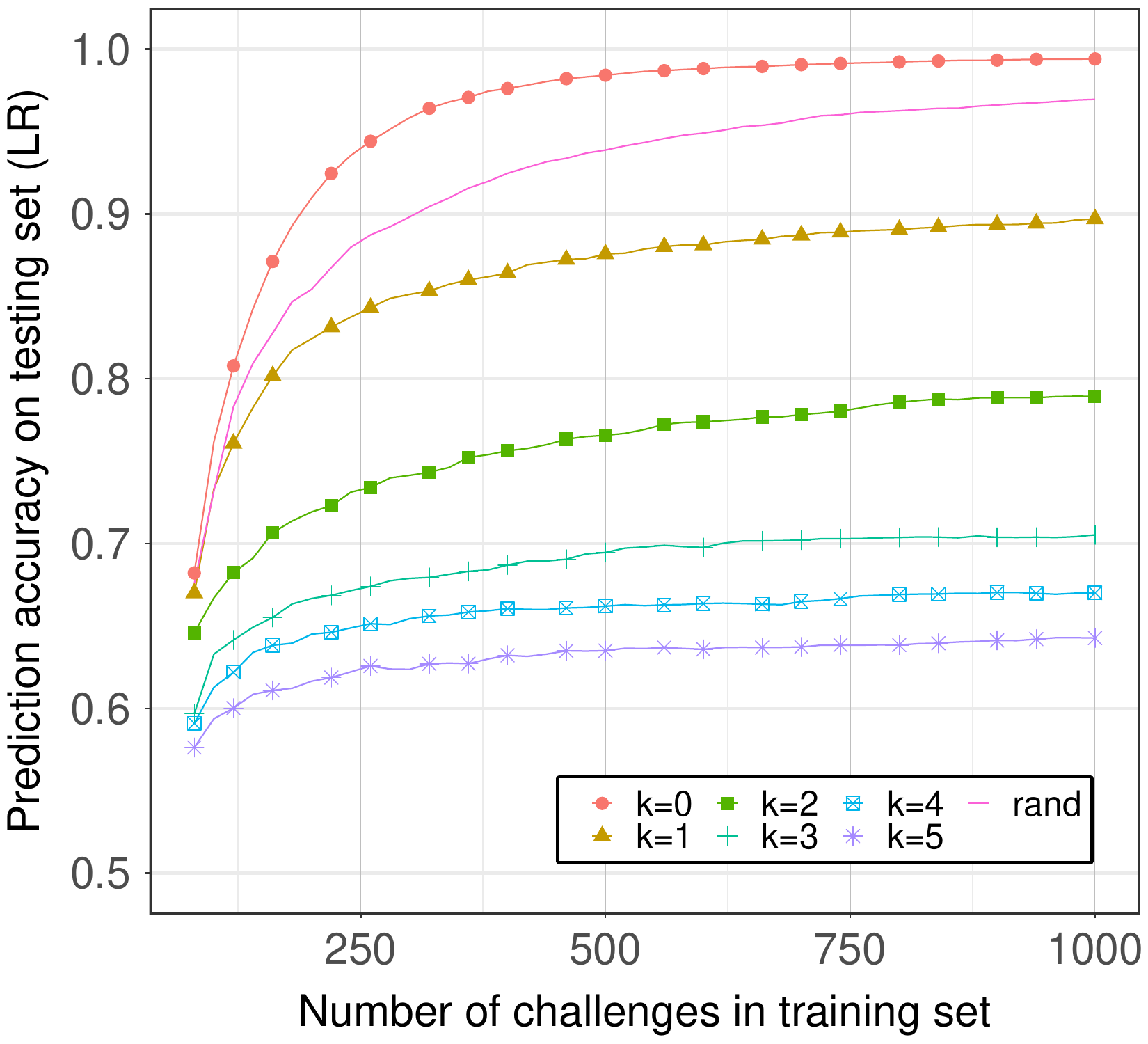}
		\caption{No noise, accuracy, LR, for a $n=64$ stage APUF}		\label{fig:LRnonoise}
\end{figure}

{\bf Comparison of SVM and LR for different $k$ values in Algorithm ``Active learning''.} 
Figures \ref{fig:SVMnonoise} (uses SVM) and \ref{fig:LRnonoise} (uses Logistic Regression or LR) compare the performance in the absence of noise. We see that SVM has much better performance than LR in general (higher accuracy for lower number of challenges); the key takeaway here is thus {\it use SVM learning for the arbiter PUF.} The blue $k=0$ line represents the best active /  fast learning for a small number of challenges, outperforming random challenge selection by quite a margin. For $k>3$ we obtain effective adversarial / slow learning results, with the SVM unable to reach much higher accuracies than about 70\%. We see a nice progression from fast to slow learning as we increase $|k|$.

\begin{figure}[h!]
	\centering	\includegraphics[width=0.45\textwidth]{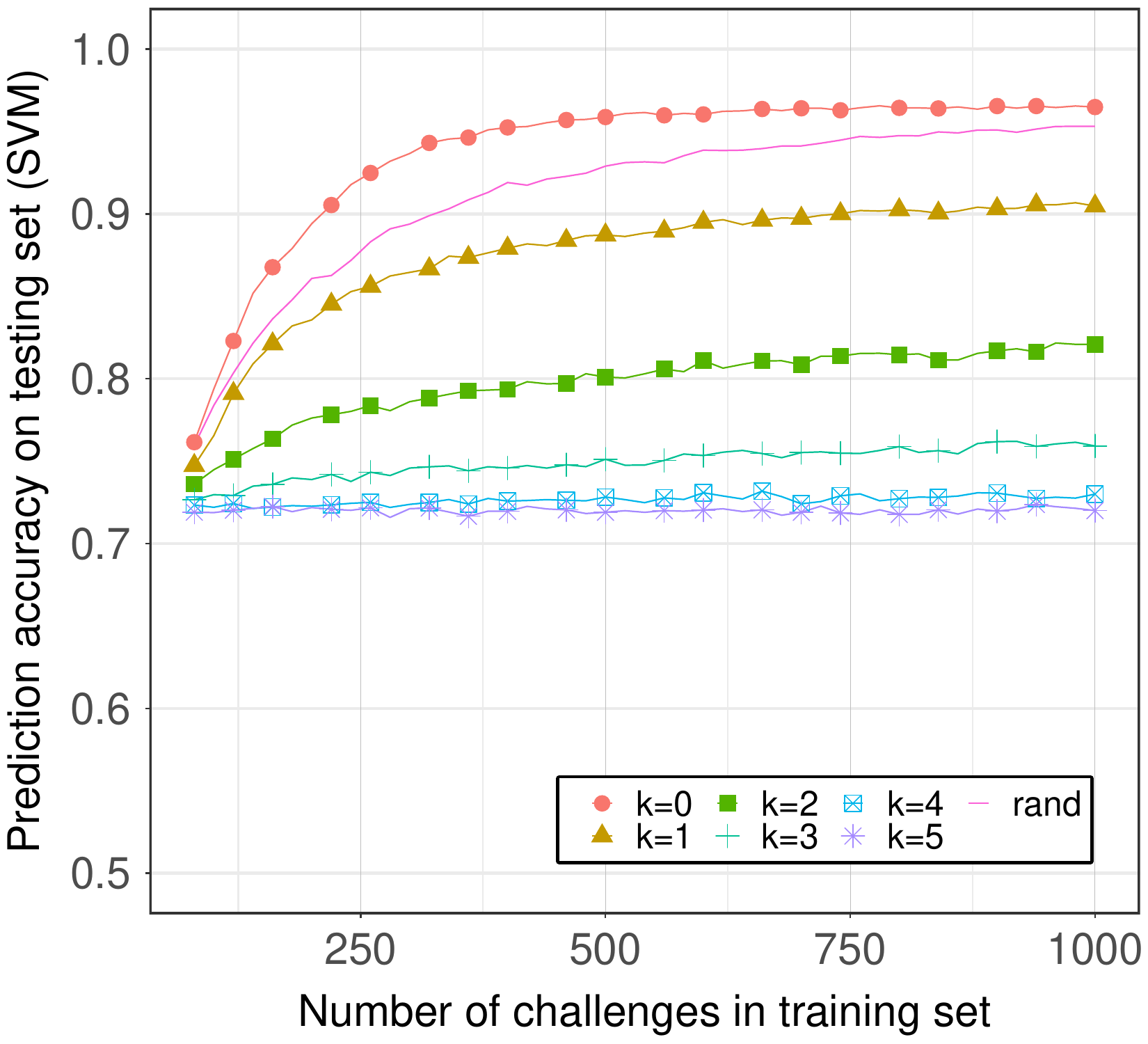}
		\caption{$3.5\%$ noise, accuracy, SVM, for a $n=64$ stage APUF}		\label{fig:SVMnoise}
\end{figure}

Figure \ref{fig:SVMnoise} shows the performance of Algorithm ``active learn''  under SVM for various $k$ in the presence of noise which flips about 3.5\% of the responses to the challenges. Here learning is slowed down a bit even for $k=0$, but still remarkably outperforms random challenges for a small number of challenges. Slow learning is  effective with $k=5$. The analogous plot for  LR under noisy samples is quite similar to Figure \ref{fig:LRnonoise} and hence omitted.

\section{Conclusion}

We have developed new methods based on challenge selection for active learning of arbiter PUFs. First, we presented a new challenge set based on the Hadamard code which yields excellent modeling accuracy when used as learning data points for a very small number of challenges (less than or equal to the number of stages of the arbiter PUF $n$). Then, we presented an active learning algorithm that iteratively constructs model-specific challenges for both active / fast learning (when a manufacturer or attacker can query the PUF at will) and adversarial / slow learning (for selecting challenges to use in an authentication protocol where an attacker  overhears the CRPs but cannot select them).  Our algorithm  uses a tunable parameter $k$ which selects a target normalized distance to the SVM hyperplane from which to select challenges. This distance roughly correlates with how ``useful'' the challenge is -- lower $|k|$ yields faster learning, higher $|k|$ yields slower learning.  
Our method does not use  a sample pool and in that sense is much more powerful than existing sample-pool-based methods (which limit the challenges available for selection).  Our experimental results validate this: for fast learning we set $k=0$  which yields the fastest/best to date known attack on an arbiter PUf, both with and without noise. For adversarial / slow learning we set $k=5$ and find a set of challenges with excellent internal recognition but poor generalization error (hovering around 70\% for SVM, LR, Neural network and AdaBoost based learning methods). We emphasize that our algorithms are remarkably simple and effective and make use of the linear threshold model of the arbiter PUF  well suited to be learned  using SVM.

\baselineskip=2pt
\bibliographystyle{IEEEtranS}
\bibliography{refs, refs-abnormal, refs-more}

\begin{thebibliography}{10}
\providecommand{\url}[1]{#1}
\csname url@rmstyle\endcsname
\providecommand{\newblock}{\relax}
\providecommand{\bibinfo}[2]{#2}
\providecommand\BIBentrySTDinterwordspacing{\spaceskip=0pt\relax}
\providecommand\BIBentryALTinterwordstretchfactor{4}
\providecommand\BIBentryALTinterwordspacing{\spaceskip=\fontdimen2\font plus
\BIBentryALTinterwordstretchfactor\fontdimen3\font minus
  \fontdimen4\font\relax}
\providecommand\BIBforeignlanguage[2]{{%
\expandafter\ifx\csname l@#1\endcsname\relax
\typeout{** WARNING: IEEEtran.bst: No hyphenation pattern has been}%
\typeout{** loaded for the language `#1'. Using the pattern for}%
\typeout{** the default language instead.}%
\else
\language=\csname l@#1\endcsname
\fi
#2}}

\bibitem{7096998}
G.~T. Becker, ``On the pitfalls of using arbiter-pufs as building blocks,''
  \emph{IEEE Trans. Comput.-Aided Des. Integr. Circuits Syst.}, vol.~34, no.~8,
  pp. 1295--1307, Aug 2015.

\bibitem{boning2000models}
A.~Chandrakasan, W.~J. Bowhill, and F.~Fox, \emph{Models of Process Variations
  in Device and Interconnect}, 2001, pp. 98--115.

\bibitem{Vapnik1995}
V.~V. Corinna~Cortes, ``Support-vector networks,'' \emph{Machine Learning},
  vol.~20, p. 273–297, 1995.

\bibitem{rep-inform}
B.~Du, Z.~Wang, L.~Zhang, L.~Zhang, W.~Liu, J.~Shen, and D.~Tao, ``Exploring
  representativeness and informativeness for active learning,'' \emph{IEEE
  Trans. on Cybern.}, vol.~47, no.~1, pp. 14--26, 2017.

\bibitem{8607165}
H.~L. França, C.~B. Prado, V.~C. Patil, and S.~Kundu, ``Defeating strong puf
  modeling attack via adverse selection of challenge-response pairs,'' in
  \emph{AsianHOST}, 2018, pp. 25--30.

\bibitem{gassend2004identification}
B.~Gassend, D.~Lim, D.~Clarke, M.~van Dijk, and S.~Devadas, ``Identification
  and authentication of integrated circuits,'' \emph{Concurrency - Practice and
  Experience}, vol.~16, pp. 1077--1098, 09 2004.

\bibitem{herder2014physical}
C.~Herder, M.-D. Yu, F.~Koushanfar, and S.~Devadas, ``Physical unclonable
  functions and applications: A tutorial,'' \emph{Proc. IEEE}, vol. 102, no.~8,
  pp. 1126--1141, 2014.

\bibitem{Kremer}
J.~Kremer, K.~Steenstrup~Pedersen, and C.~Igel, ``Active learning with support
  vector machines,'' \emph{WIREs Data Mining and Knowledge Discovery}, vol.~4,
  no.~4, pp. 313--326, 2014.

\bibitem{adaboost_tree:2013}
B.~Kégl, ``The return of adaboost.mh: multi-class hamming trees,'' 2013.

\bibitem{rioul2016entropy}
O.~Rioul, P.~Sol{\'e}, S.~Guilley, and J.-L. Danger, ``On the entropy of
  physically unclonable functions,'' in \emph{ISIT}.\hskip 1em plus 0.5em minus
  0.4em\relax IEEE, 2016, pp. 2928--2932.

\bibitem{ruhrmair2010modeling}
U.~R{\"u}hrmair, F.~Sehnke, J.~S{\"o}lter, G.~Dror, S.~Devadas, and
  J.~Schmidhuber, ``Modeling attacks on physical unclonable functionso,'' in
  \emph{CCS}.\hskip 1em plus 0.5em minus 0.4em\relax ACM, 2010, pp. 237--249.

\bibitem{ruhrmair2013puf}
U.~R{\"u}hrmair, J.~S{\"o}lter, F.~Sehnke, X.~Xu, A.~Mahmoud, V.~Stoyanova,
  G.~Dror, J.~Schmidhuber, W.~Burleson, and S.~Devadas, ``Puf modeling attacks
  on simulated and silicon data,'' \emph{IEEE Trans. on Info. Foren. and Sec.},
  vol.~8, no.~11, pp. 1876--1891, 2013.

\bibitem{us-TCAD}
{same authors, kept anonymous for peer revision}, ``{Understanding
  Challenge-Response-Pair Correlations in Arbiter PUFs}.''

\bibitem{Sylvester1867}
J.~Sylvester, ``Thoughts on inverse orthogonal matrices, simultaneous sign
  successions, and tessellated pavements in two or more colours, with
  applications to newton's rule, ornamental tile-work, and the theory of
  numbers,'' \emph{Philosophical Magazine}, vol.~34, pp. 461--475, 1867.

\bibitem{active-PUF}
Y.~Wen and Y.~Lao, ``Puf modeling attack using active learning,'' \emph{ISCAS},
  pp. 1--5, 2018.

\bibitem{SVM-wiki}
\BIBentryALTinterwordspacing
Wikipedia, ``Support vector machines.'' [Online]. Available:
  \url{https://en.wikipedia.org/wiki/Support\_vector\_machine}
\BIBentrySTDinterwordspacing

\end{thebibliography}

\end{document}